\documentclass{article}
\usepackage{spconf,amsmath,graphicx}
\usepackage{amssymb,amsfonts,bm}
\usepackage[export]{adjustbox} 
\usepackage[ruled,vlined]{algorithm2e}
\usepackage{hyperref}
\usepackage[caption=false]{subfig}
\usepackage{color,soul}
\usepackage[table,xcdraw]{xcolor}
\usepackage{multirow}
\usepackage{cite}
\usepackage{siunitx}
\usepackage[utf8]{inputenc}
\usepackage[strings]{underscore} 
\usepackage[english]{babel} 
\usepackage{duckuments} 

\usepackage{pifont}
\newcommand{\cmark}{\text{\ding{51}}} 
\newcommand{\xmark}{\text{\ding{55}}} 

\usepackage[para]{threeparttable}

\usepackage{floatrow}
\usepackage{wrapfig}

\usepackage[subtle,tracking=normal]{savetrees}
\usepackage{setspace}

\usepackage[inline]{enumitem}
\newlist{inlinelist}{enumerate*}{1} 
\setlist[inlinelist]{label=(\roman*)}

\setlist[itemize]{nosep} 

\newcommand{\dynbox}[2][c]{\begin{tabular}{@{}#1@{}}#2\end{tabular}}

\DeclareMathOperator*{\argmin}{arg\,min}

\title{cRedAnno+: Annotation Exploitation in Self-Explanatory \\ Lung Nodule Diagnosis}
\name{\dynbox{Jiahao Lu$^{1,2}$ \quad  Chong Yin$^{3}$ \quad 
Kenny Erleben$^{1}$ \quad Michael Bachmann Nielsen$^{2}$ \quad Sune Darkner$^{1}$}}
\address{$^{1}$Department of Computer Science, University of Copenhagen, Denmark \\
    $^{2}$Department of Diagnostic Radiology, Rigshospitalet, Copenhagen University Hospital, Denmark\\
    $^{3}$Department of Computer Science, Hong Kong Baptist University, China}
\begin{document}
%
\maketitle
\begin{abstract}
Recently, attempts have been made to reduce annotation requirements in feature-based self-explanatory models for lung nodule diagnosis. As a representative, cRedAnno achieves competitive performance with considerably reduced annotation needs by introducing self-supervised contrastive learning to do unsupervised feature extraction.
However, it exhibits unstable performance under scarce annotation conditions. 
To improve the accuracy and robustness of cRedAnno, we propose an annotation exploitation mechanism by conducting semi-supervised active learning with sparse seeding and training quenching in the learned semantically meaningful reasoning space to jointly utilise the extracted features, annotations, and unlabelled data.
The proposed approach achieves comparable or even higher malignancy prediction accuracy with 10x fewer annotations, meanwhile showing better robustness and nodule attribute prediction accuracy under the condition of $1\%$ annotations. 
Our complete code is open-source available: \url{https://github.com/diku-dk/credanno}.
\end{abstract}

\section{Introduction}
\label{sec:intro}

Effective lung cancer screening requires accurate characterisation of pulmonary nodules in CT images \cite{vlahosLungCancerScreening2018}. 
Amongst recent efforts in explainable AI, post-hoc approaches that attempt to interpret well-preforming ``black boxes" \cite{liuMultiTaskDeepModel2020} are not deemed trustworthy enough in clinical practice \cite{rudinStopExplainingBlack2019, vanderveldenExplainableArtificialIntelligence2022}. 
In contrast, feature-based self-explanatory methods are trained to use a set of well-known human-understandable concepts to explain and derive their decisions \cite{shenInterpretableDeepHierarchical2019, lalondeEncodingVisualAttributes2020}. 
Although such semantic matching towards clinical knowledge is especially valuable in medical applications \cite{salahuddinTransparencyDeepNeural2022}, the additional annotation requirements for features still limit the applicability of this approach.

\begin{figure}[tbp]
    \centering
    \includegraphics[width=0.8\textwidth]{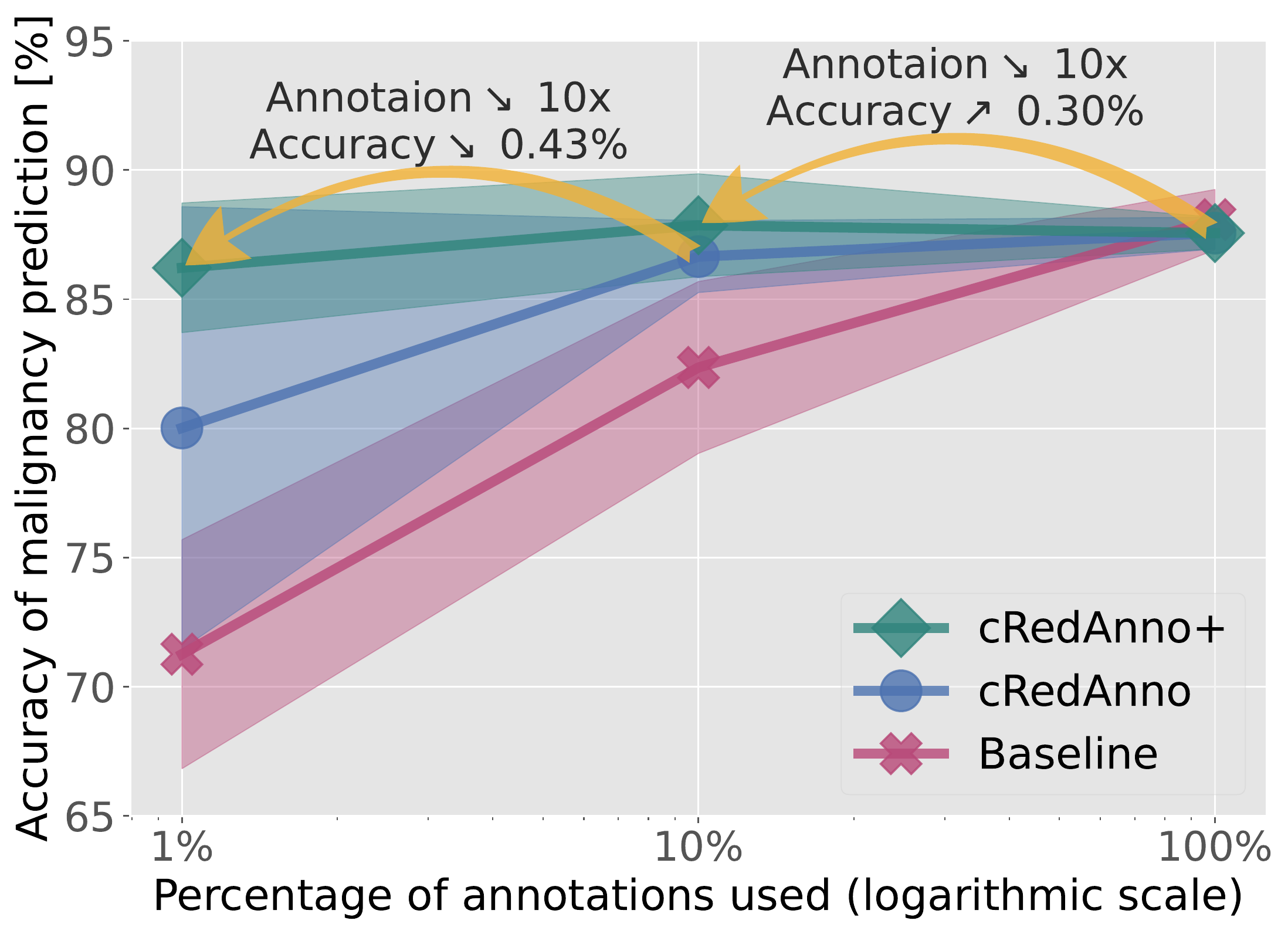}

    \caption{\textbf{Comparison between \textcolor[HTML]{2F847C}{cRedAnno+} and \textcolor[HTML]{4C72B0}{cRedAnno}}, in terms of nodule malignancy prediction accuracy and annotation cost.
    cRedAnno+ achieves comparable or even higher accuracy with 10x fewer annotations, meanwhile being more robust.
    }
    \label{fig:intro_compare}
    \vspace{-0.7ex}
\end{figure}

The recently proposed cRedAnno \cite{luReducingAnnotationNeed2022} addresses this problem by introducing self-supervised contrastive learning \cite{caronEmergingPropertiesSelfSupervised2021} to alleviate the burden of learning most parameters from annotations. 
Albeit cRedAnno \cite{luReducingAnnotationNeed2022} achieves competitive performance using hundreds of nodule samples and considerably reduced annotations, its robustness awaits improvement when labelled samples are extremely scarce (Fig.~\ref{fig:intro_compare}), whereas the unlabelled data are not adequately utilised yet \cite{simeoniRethinkingDeepActive2021}.

\begin{figure}[tbp]
    \centering
    \includegraphics[width=\textwidth]{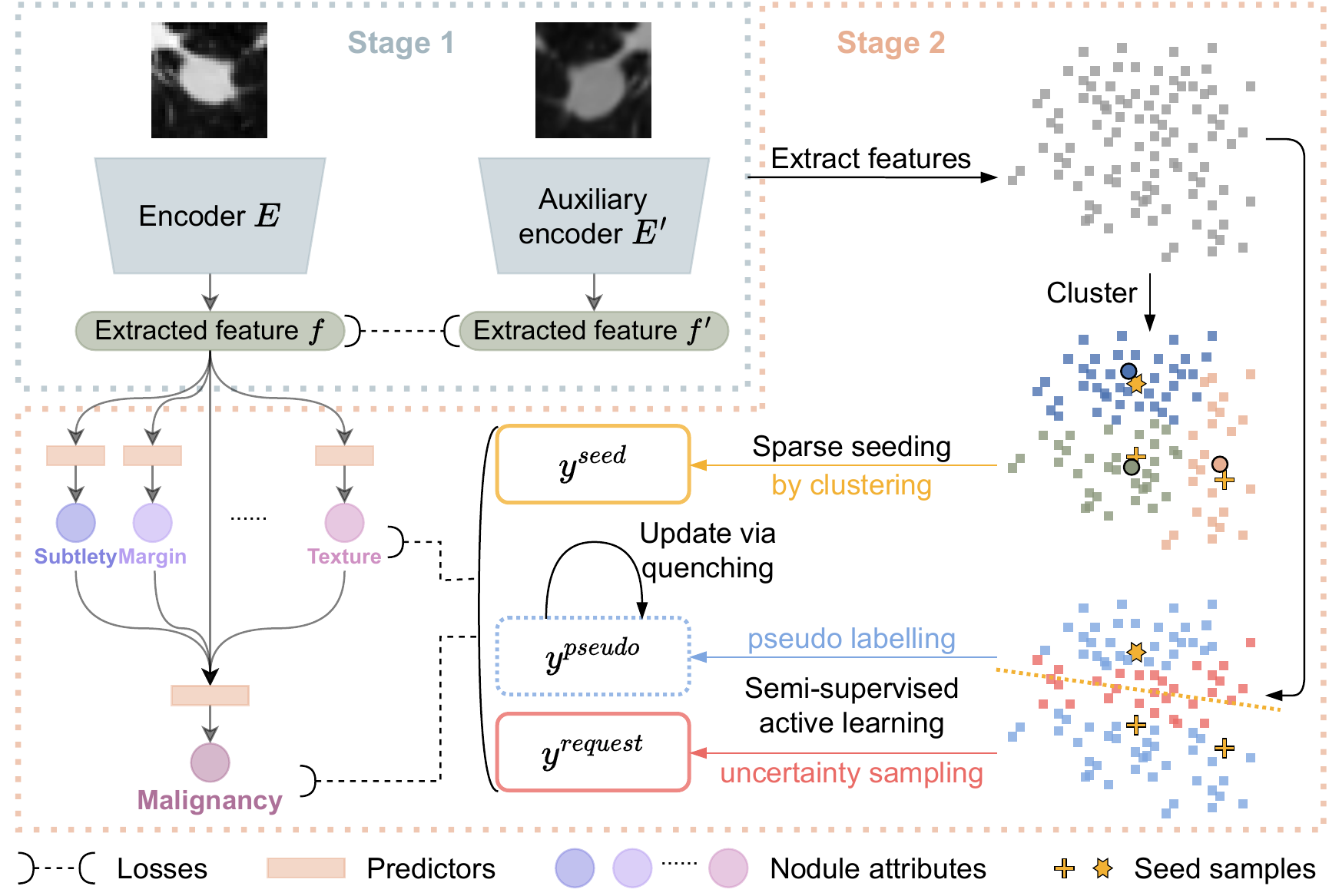}

    \caption{\textbf{Concept illustration.}
    The proposed annotation exploitation mechanism conducts semi-supervised active learning with sparse seeding and training quenching in the semantically meaningful space learned in Stage 1, to jointly exploit the extracted features, annotations, and unlabelled data.
    }
    \vspace{-0.7ex}
    \label{fig:overview}
\end{figure}

This paper aims to improve the prediction accuracy and robustness of cRedAnno \cite{luReducingAnnotationNeed2022} under scarce annotation conditions. We address this by conducting semi-supervised active learning\cite{tamkinActiveLearningHelps2022} in the learned latent space that complies with radiologists' reasoning for nodule malignancy. More specifically, we propose an efficient annotation exploitation mechanism, composing seeding by clustering\cite{huGoodStartUsing2010}, uncertainty sampling\cite{settlesUncertaintySampling2012}, and pseudo labelling\cite{wangCostEffectiveActiveLearning2017} to jointly utilise the extracted features, annotations, and unlabelled data, facilitated by a quenching technique to update the pseudo labels and reinitialise the weights of predictors, as shwon in Fig.~\ref{fig:overview}.


Compared with cRedAnno \cite{luReducingAnnotationNeed2022}, the proposed cRedAnno+ achieves comparable or even higher malignancy prediction accuracy with $10$x fewer annotations whilst reaching simultaneously above $90\%$ mean accuracy in predicting all nodule attributes, meanwhile being more robust under the condition of $1\%$ annotations (Fig.~\ref{fig:intro_compare}).

\section{Method}
\label{sec:method}

As illustrated in Fig.~\ref{fig:overview}, the proposed approach enriches cRedAnno \cite{luReducingAnnotationNeed2022} by replacing its second-stage supervised predictor training using random samples with an annotation exploitation mechanism. Therefore, here we only outline what is inherited from cRedAnno, whilst detailing the proposed approach and focusing on their differences.

\subsection{Recapitulation of cRedAnno}
\label{sec:method_recap}
The original cRedAnno\cite{luReducingAnnotationNeed2022} uses two-stage training, replacing the end-to-end training paradigm in previous works.
In Stage 1, the majority of parameters are trained using self-supervised contrastive learning \cite{caronEmergingPropertiesSelfSupervised2021} as an encoder to map the input images to a latent space that complies with radiologists' reasoning for nodule malignancy. 
In Stage 2, a small random portion of labelled samples is used to train a hierarchical set of predictors $\{G_\text{cls}, G_\text{exp}\}$, including a predictor $G_\text{exp}^{(i)}$ for each nodule attribute $i$, and a malignancy predictor $G_\text{cls}$ whose input is the concatenation of extracted features $f$ and the predicted human-understandable nodule attributes.

\textbf{Limitations of cRedAnno \cite{luReducingAnnotationNeed2022}.} The high prediction accuracy exhibited by cRedAnno\cite{luReducingAnnotationNeed2022} can attribute to the extracted features being highly separable in the learned space. Nevertheless, when only very few annotations are available, the random selection is likely biased and even risks not covering enough label space, which may lead to severe performance degradation. Furthermore, for training the predictors on the self-supervised model, randomly selected annotations are not necessarily informative enough\cite{tamkinActiveLearningHelps2022}.
In addition, the unlabelled data are not adequately used in training the predictors.

\begin{figure}[tbp]
    \centering
    \subfloat[Malignancy]{
        \includegraphics[width=0.229\textwidth]{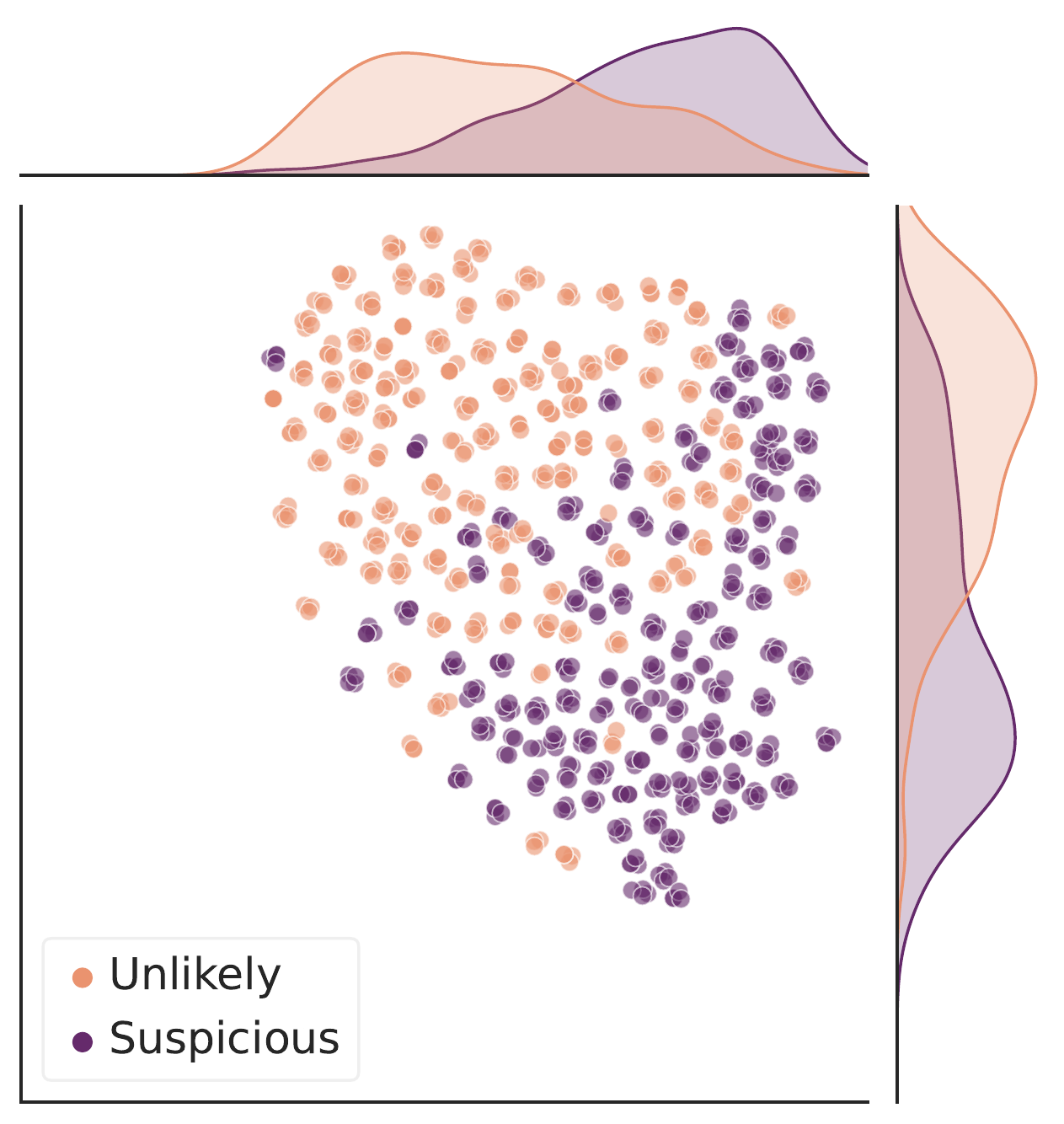}
        \label{fig:tsne_mal}
        }\hfil
    \subfloat[Subtlety]{
        \includegraphics[width=0.229\textwidth]{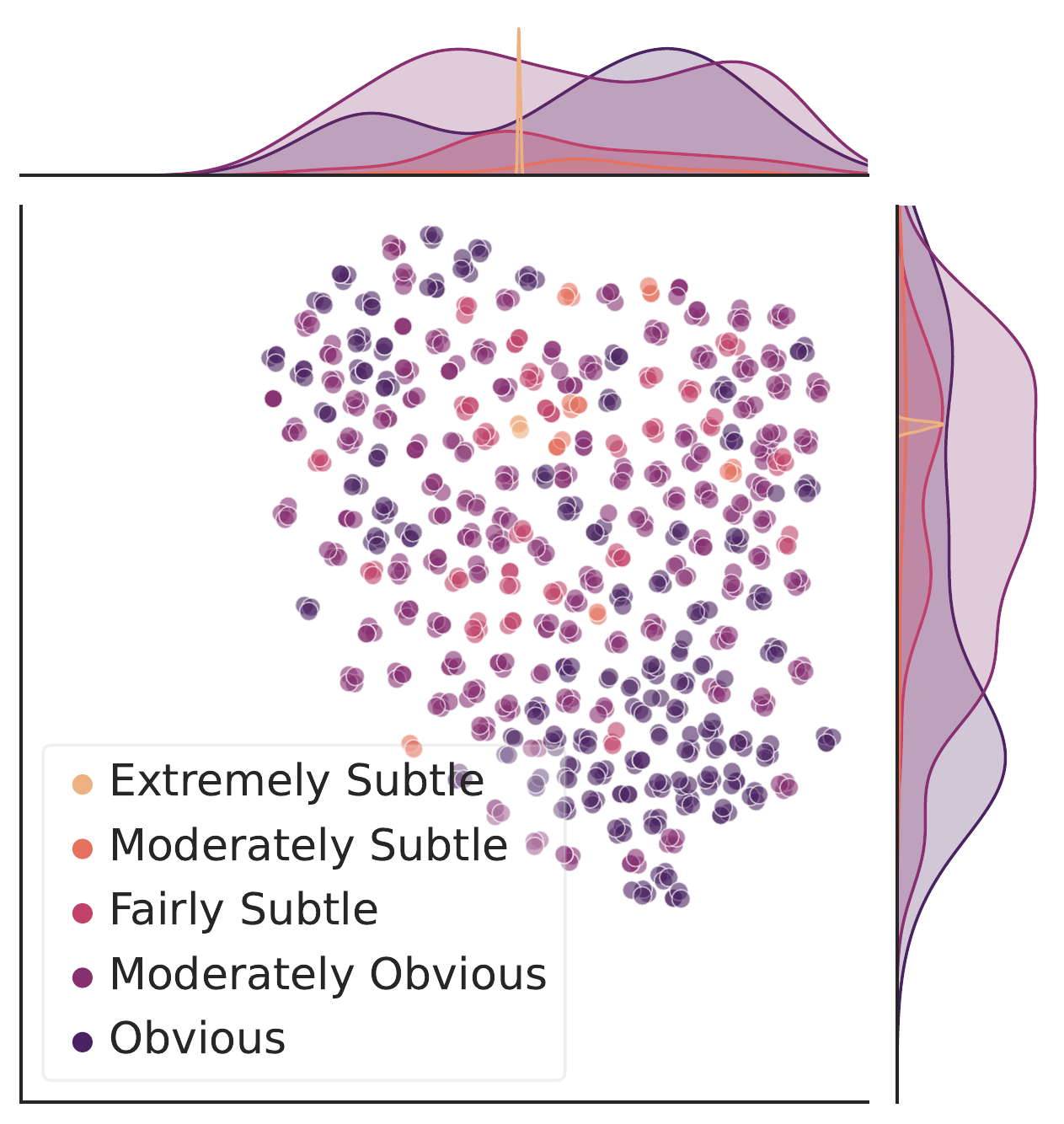}
        \label{fig:tsne_sub}
        }\hfil
    \subfloat[Calcification]{
        \includegraphics[width=0.229\textwidth]{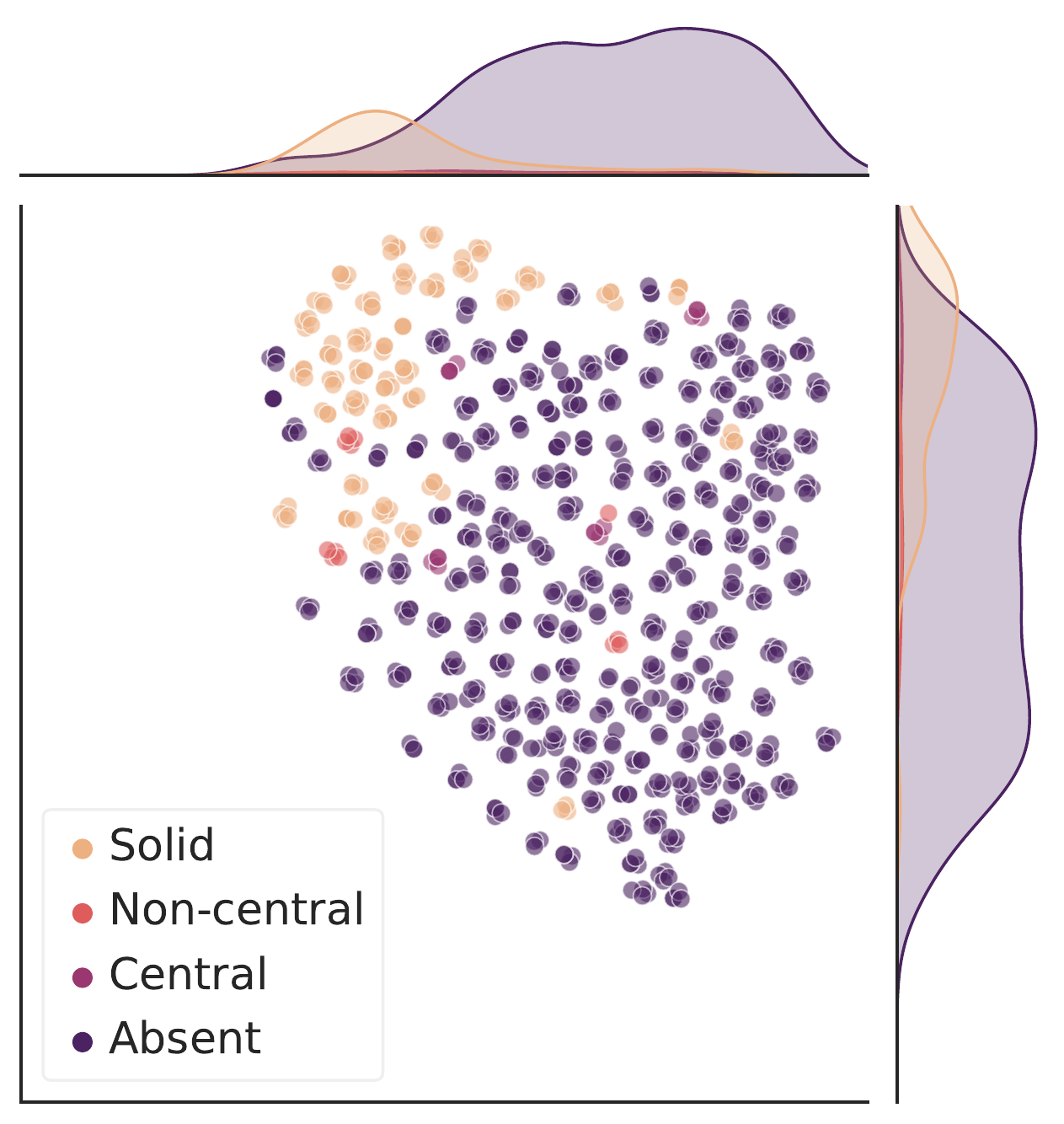}
        \label{fig:tsne_cal}
        }\hfil
    \subfloat[Sphericity]{
        \includegraphics[width=0.229\textwidth]{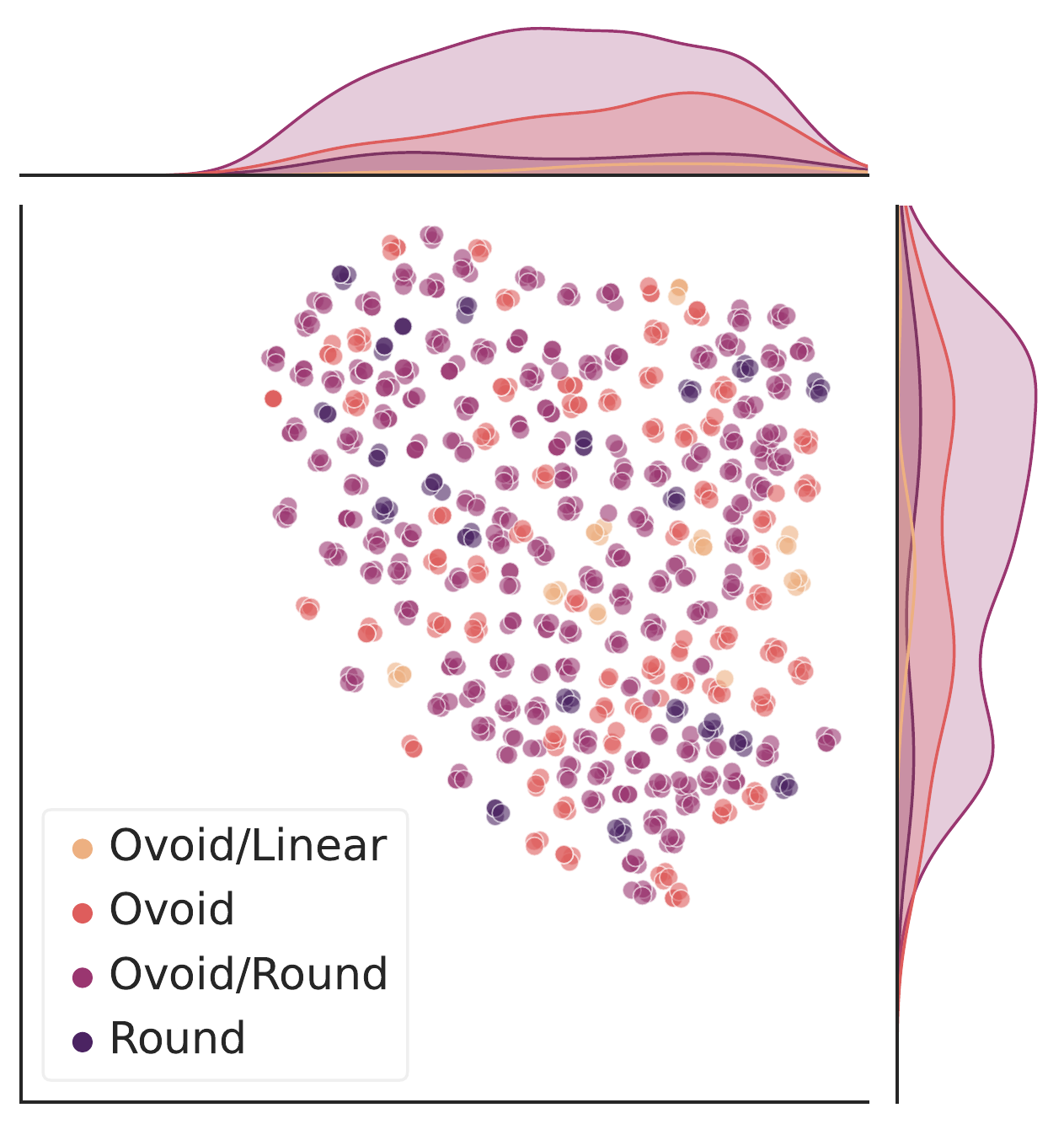}
        \label{fig:tsne_sph}
        }\\[-1ex]

    \subfloat[Margin]{
        \includegraphics[width=0.229\textwidth]{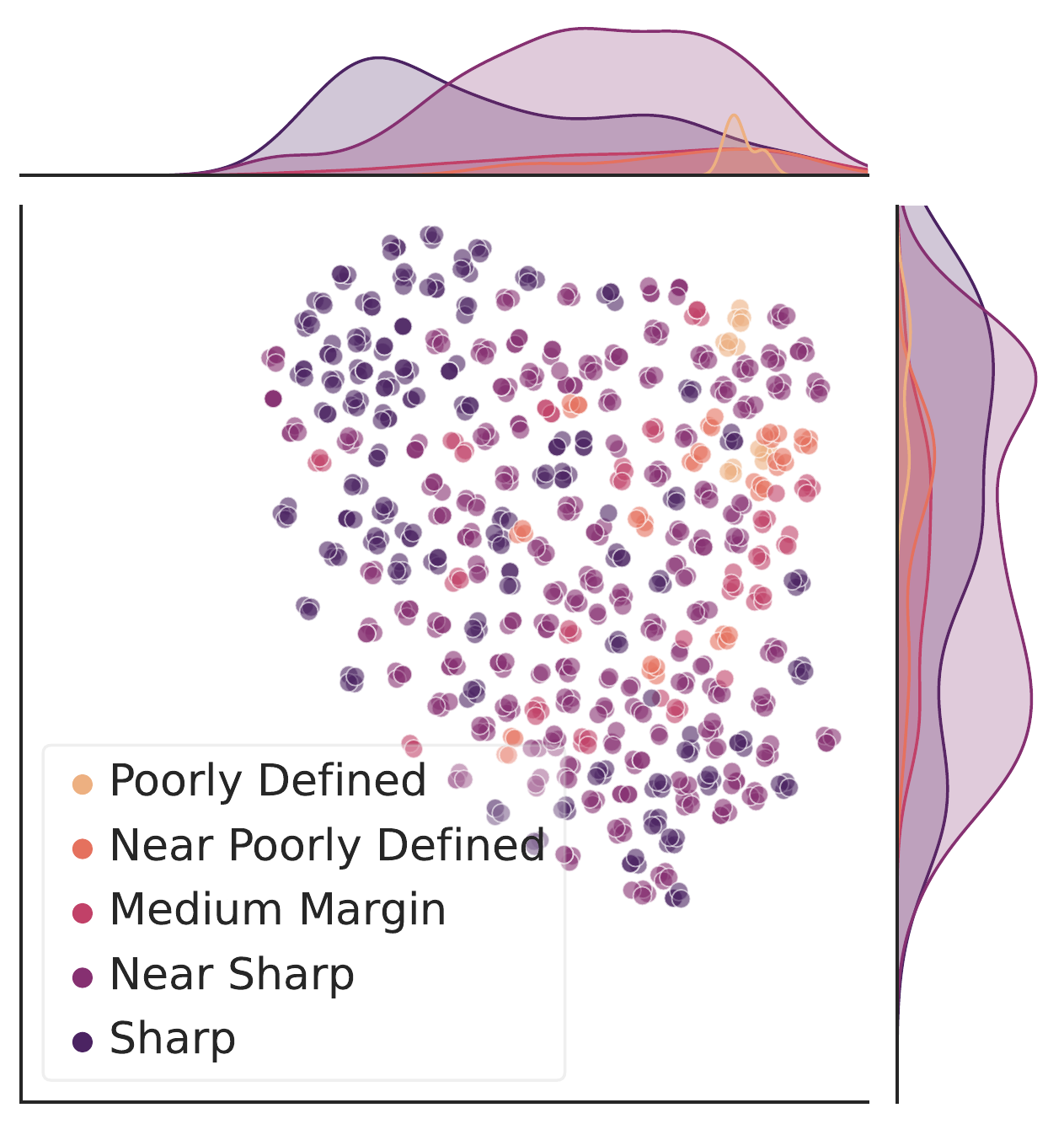}
        \label{fig:tsne_mar}
        }\hfil
    \subfloat[Lobulation]{
        \includegraphics[width=0.229\textwidth]{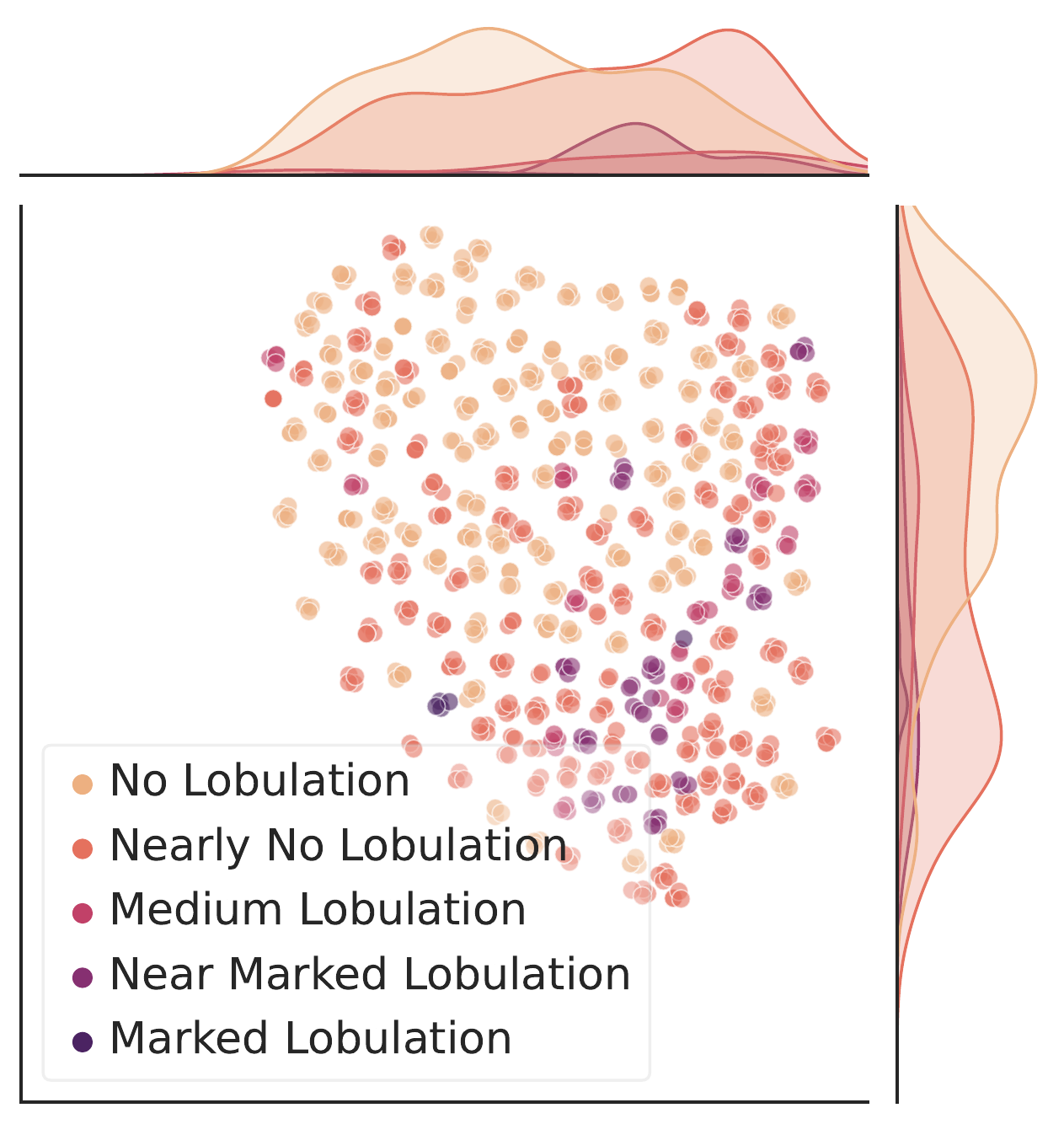}
        \label{fig:tsne_lob}
        }\hfil
    \subfloat[Spiculation]{
        \includegraphics[width=0.229\textwidth]{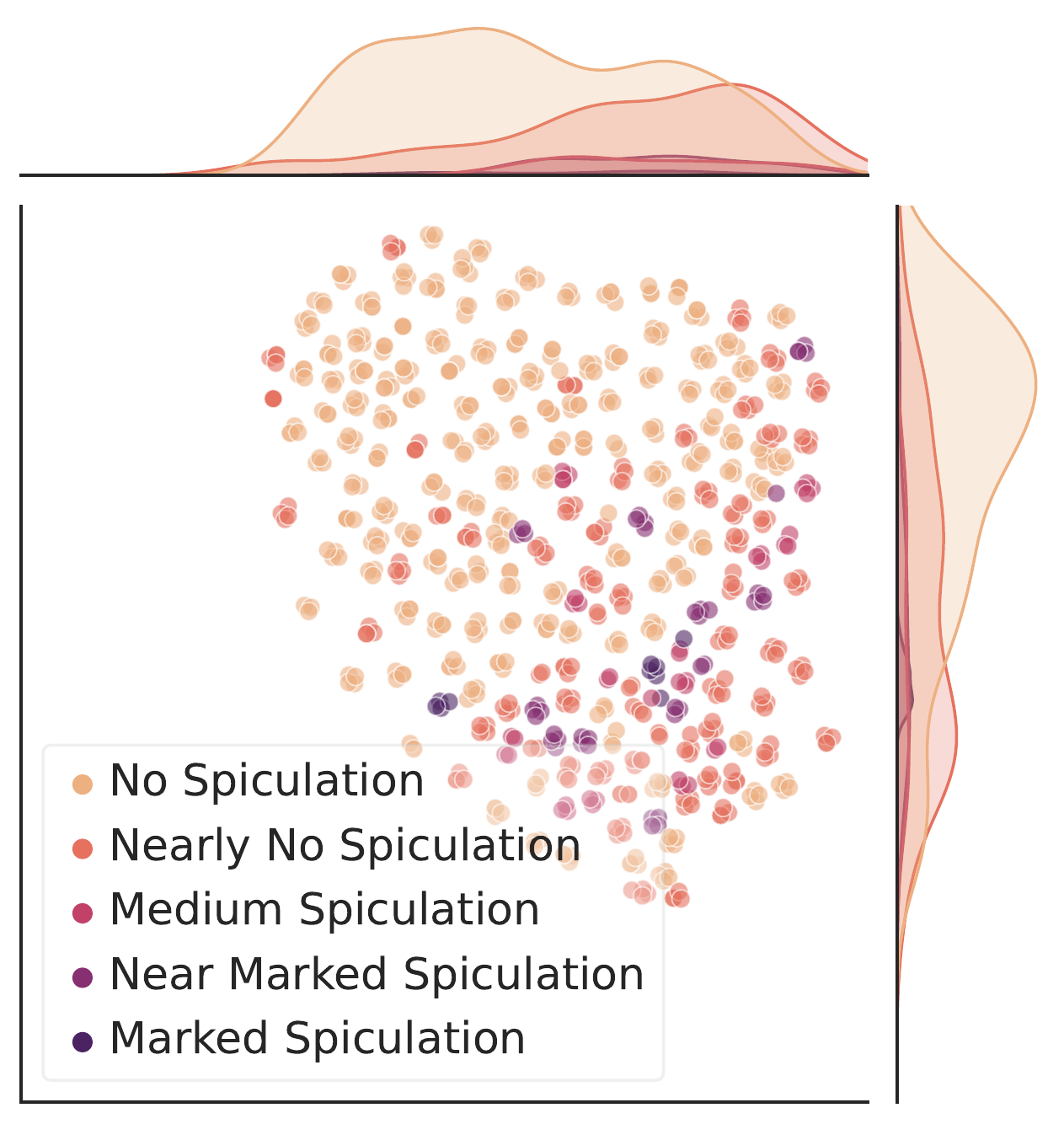}
        \label{fig:tsne_spi}
        }\hfil
    \subfloat[Texture]{
        \includegraphics[width=0.229\textwidth]{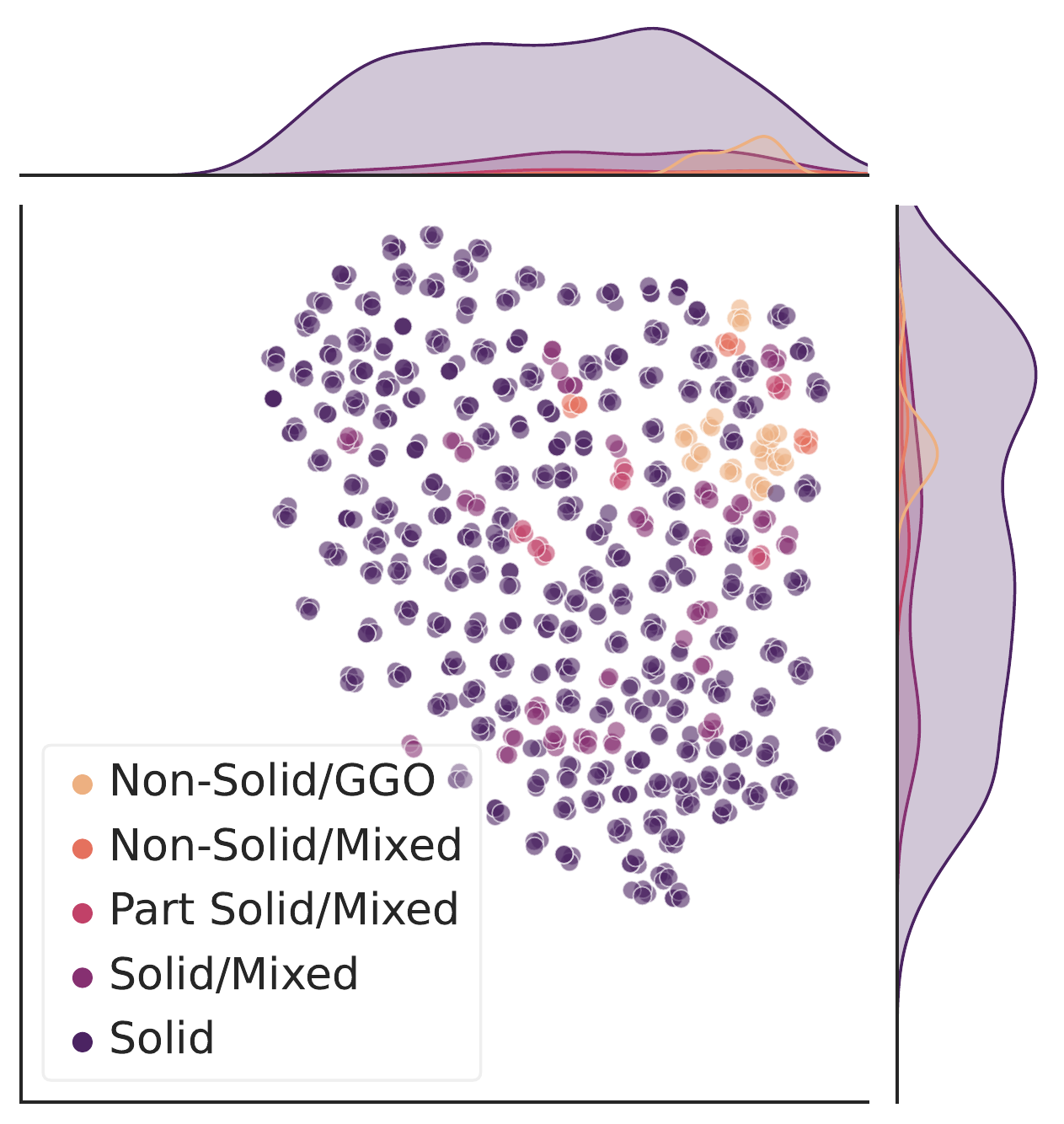}
        \label{fig:tsne_tex}
        }\\[-0.7ex]

    \caption{\textbf{t-SNE visualisation of features extracted from testing images.}
    Data points are coloured using ground truth annotations.
    Whilst malignancy shows highly separable and semantically correlates with the nodule attributes, a random selection of few samples may risk not covering enough label space.
    }
    \label{fig:res_tsne}
\end{figure}

\begin{table*}[t]
    \centering
    \caption{\textbf{Prediction accuracy ($\%$) of nodule attributes and malignancy.}
    The best in each column is \textbf{bolded} for full/partial annotation respectively.
    Dashes (-) denote values not reported by the compared methods. 
    Results of our proposed \colorbox[HTML]{E2EFD9}{cRedAnno+} and the previous \colorbox[HTML]{DFE7F3}{cRedAnno}\cite{luReducingAnnotationNeed2022} are highlighted.
    Observe that with $1\%$ annotations, cRedAnno+ reaches competitive accuracy in malignancy prediction and over $90\%$ accuracy simultaneously in predicting all nodule attributes, meanwhile using the fewest nodules and no additional information.
    }
    \label{tab:res}
    \resizebox{\textwidth}{!}{%
    \begin{threeparttable}
        \begin{tabular}{lcccccccccc}
            \hline
             & \multicolumn{7}{c}{\textbf{Nodule attributes}} & \multicolumn{1}{l}{} & \multicolumn{1}{l}{} & \multicolumn{1}{l}{} \\ \cline{2-8}
             & \textbf{Sub} & \textbf{Cal} & \textbf{Sph} & \textbf{Mar} & \textbf{Lob} & \textbf{Spi} & \textbf{Tex} & \multicolumn{1}{l}{\multirow{-2}{*}{\textbf{Malignancy}}} & \multicolumn{1}{l}{\multirow{-2}{*}{\textbf{\#nodules}}} & \multicolumn{1}{l}{\multirow{-2}{*}{\begin{tabular}[c]{@{}c@{}}\textbf{No additional} \\ \textbf{information}\end{tabular}}} \\ \hline

            \multicolumn{11}{l}{Full annotation} \\
            \textbf{HSCNN}\cite{shenInterpretableDeepHierarchical2019} & 71.90 & 90.80 & 55.20 & 72.50 & - & - & 83.40 & 84.20 & 4252 & \xmark\tnote{c} \\
            \textbf{X-Caps}\cite{lalondeEncodingVisualAttributes2020} & 90.39 & - & 85.44 & 84.14 & 70.69 & 75.23 & 93.10 & 86.39 & 1149 & \cmark \\
            \textbf{MSN-JCN}\cite{chenEndtoEndMultiTaskLearning2021} & 70.77 & 94.07 & 68.63 & 78.88 & \textbf{94.75} & 93.75 & 89.00 & 87.07 & 2616 & \xmark\tnote{d} \\
            \textbf{MTMR}\cite{liuMultiTaskDeepModel2020} & - & - & - & - & - & - & - & \textbf{93.50} & 1422 & \xmark\tnote{e} \\
            \rowcolor[HTML]{E2EFD9} 
            \textbf{cRedAnno+} & \textbf{96.32$\pm$0.61} & \textbf{95.88$\pm$0.15} & \textbf{97.23$\pm$0.20} & \textbf{96.23$\pm$0.23} & 93.93$\pm$0.87 & \textbf{94.06$\pm$0.60} & \textbf{97.01$\pm$0.26} & 87.56$\pm$0.61 & \textbf{730} & \cmark \\ \hline

            \multicolumn{11}{l}{Partial annotation} \\
            \textbf{WeakSup}\cite{joshiLungNoduleMalignancy2021} \textbf{(1:5\tnote{a} )} & 43.10 & 63.90 & 42.40 & 58.50 & 40.60 & 38.70 & 51.20 & 82.40 &  &  \\
            \textbf{WeakSup}\cite{joshiLungNoduleMalignancy2021} \textbf{(1:3\tnote{a} )} & 66.80 & 91.50 & 66.40 & 79.60 & 74.30 & 81.40 & 82.20 & \textbf{89.10} & \multirow{-2}{*}{2558} & \multirow{-2}{*}{\xmark\tnote{f}} \\
            \rowcolor[HTML]{DFE7F3} 
            \textbf{cRedAnno (10\%\tnote{b} )} & 96.06$\pm$2.02 & \textbf{93.76$\pm$0.85} & \textbf{95.97$\pm$0.69} & \textbf{94.37$\pm$0.79} & 93.06$\pm$0.27 & 93.15$\pm$0.33 & \textbf{95.49$\pm$0.85} & 86.65$\pm$1.39 & \cellcolor[HTML]{E2EFD9} & \cellcolor[HTML]{E2EFD9} \\
            \rowcolor[HTML]{E2EFD9} 
            \textbf{cRedAnno+ (10\%\tnote{b} )} & \textbf{96.23$\pm$0.45} & 92.72$\pm$1.66 & 95.71$\pm$0.47 & 90.03$\pm$3.68 & \textbf{93.89$\pm$1.41} & \textbf{93.67$\pm$0.64} & 92.41$\pm$1.05 & 87.86$\pm$1.99 & \cellcolor[HTML]{E2EFD9} & \cellcolor[HTML]{E2EFD9} \\
            \rowcolor[HTML]{DFE7F3} 
            \textbf{cRedAnno (1\%\tnote{b} )} & 93.98$\pm$2.09 & 89.68$\pm$3.52 & 94.02$\pm$2.30 & 91.94$\pm$1.17 & 91.03$\pm$1.72 & 90.81$\pm$1.56 & 93.63$\pm$0.47 & 80.02$\pm$8.56 & \cellcolor[HTML]{E2EFD9} & \cellcolor[HTML]{E2EFD9} \\
            \rowcolor[HTML]{E2EFD9} 
            \textbf{cRedAnno+ (1\%\tnote{b} )} & 95.84$\pm$0.34 & 92.67$\pm$1.24 & \textbf{95.97$\pm$0.45} & 91.03$\pm$4.65 & 93.54$\pm$0.87 & 92.72$\pm$1.19 & 92.67$\pm$1.50 & 86.22$\pm$2.51 & \multirow{-4}{*}{\cellcolor[HTML]{E2EFD9}\textbf{730}} & \multirow{-4}{*}{\cellcolor[HTML]{E2EFD9}\cmark} \\ \hline
        \end{tabular}%
        \begin{tablenotes}
            \item[a] $1:N$ indicates that $\frac{1}{1+N}$ of training samples have annotations on nodule attributes. (All samples have malignancy annotations.)
            \item[b] The proportion of training samples that have annotations on nodule attributes and malignancy.
            \item[c] 3D volume data are used.
            \item[d] Segmentation masks and nodule diameter information are used. Two other traditional methods are used to assist training.
            \item[e] All 2D slices in 3D volumes are used.
            \item[f] Multi-scale 3D volume data are used.
        \end{tablenotes}
    \end{threeparttable}
    }
    \vspace{-0.7ex}
\end{table*}

\subsection{Annotation exploitation mechanism}
\label{sec:method_annoexp}
To jointly utilise the extracted features, annotations, and unlabelled data, we propose an annotation exploitation mechanism  (Fig.~\ref{fig:overview}) integrating the following components to address the aforementioned limitations.

\textbf{Sparse seeding.} 
To mitigate potential bias and randomness, we select seed samples by clustering\cite{huGoodStartUsing2010} the extracted features in the learned space, which was underutilised in the previous cRedAnno\cite{luReducingAnnotationNeed2022}. The extracted features are clustered into $n$ clusters, where $n$ equals the number of seed samples to select. Then the sample closest to each cluster centroid (based on cosine similarity) is selected as $f^{seed}$, whose annotations $y^{seed} = \{y^{seed}_\text{cls}, y^{seed}_\text{exp}\}$ are used to train the predictors from the initial status $\text{st}_0$ to the seeded status $\text{st}_1$: 
\begin{equation}
    G^{\text{st}_1} = \argmin_{\{G_\text{cls}, G_\text{exp}\}} \mathcal{L} \Bigl( y^{seed}, \texttt{softmax}\left( G^{\text{st}_0}(f^{seed}) \right) \Bigr),
\end{equation}
where $\mathcal{L}$ denotes the cross-entropy loss.

\textbf{Semi-supervised active learning.}
Semi-supervised learning and active learning are conducted simultaneously\cite{wangCostEffectiveActiveLearning2017} to exploit the available data. We adopt the classic yet effective uncertainty sampling by least confidence as the acquisition strategy \cite{settlesUncertaintySampling2012} to request annotations $y^{request}$ for the uncertain/informative samples $f^{request}$. Concurrently, other samples with relatively high confidence are assigned with pseudo annotations $y^{pseudo(\text{st}_t)}$ by the prediction of $G^{\text{st}_t}$.

\textbf{Quenching.}
To facilitate training under the restrictions of limited samples and complex annotation space, we propose "quenching" as a training technique. 
Similar to Curriculum Pseudo Labelling\cite{cascante-bonillaCurriculumLabelingRevisiting2021, zhangFlexMatchBoostingSemiSupervised2021}, at a certain status $\text{st}_t$ since $\text{st}_1$, the predictor weights are reinitialised to $G^{\text{st}_0}$ to avoid potential confirmation bias\cite{arazoPseudoLabelingConfirmationBias2020}. Meanwhile, the pseudo annotations are updated to the current prediction results: 
\begin{equation}
\begin{split}
    G^{\text{st}_{t+1}} &= \argmin_{\{G_\text{cls}, G_\text{exp}\}} \mathcal{L} \Bigl( \{y^{request}, y^{pseudo(\text{st}_t)}\}, \\
    &\texttt{softmax}\left( G^{\text{st}_0}\left( \{f^{request}, f^{pseudo}\}\right) \right) \Bigl),
\end{split}
\end{equation}
to preserve the learned information and resume training.

\section{Experimental results}
\label{sec:experiments}

\textbf{Data pre-processing.}
The data pre-processing remains the same as in cRedAnno \cite{luReducingAnnotationNeed2022}, following the common pre-processing procedure of the LIDC dataset\cite{armatoLungImageDatabase2011} summarised in \cite{baltatzisPitfallsSampleSelection2021}, resulting in $276/242$ benign/malignant nodules for training and $108/104$ benign/malignant nodules for testing.

\noindent
\textbf{Training settings.}
Our training settings in Stage 1 remain the same as in cRedAnno \cite{luReducingAnnotationNeed2022}. In Stage 2, K-means are used for clustering to select $1\%$ annotation as seed samples. The predictors $G_\text{exp}^{(i)}$ and $G_\text{cls}$, each consisting of one linear layer, are first jointly trained using the seed samples for $100$ epochs with SGD optimisers with momentum $0.9$ and batch size $128$. The learning rate follows a cosine scheduler with initial value $0.00025$. 
After reaching the seeded status $G^{\text{st}_1}$, the predictors and optimisers are quenched for the first time. The training then resumes using the requested and dynamic pseudo annotations for $50$ more epochs, where quenching happens every $10$ epochs.

\begin{figure}[tbp]
    \vspace{-1ex}
    \centering
    \subfloat[Full annotation]{
        \includegraphics[width=0.47\textwidth]{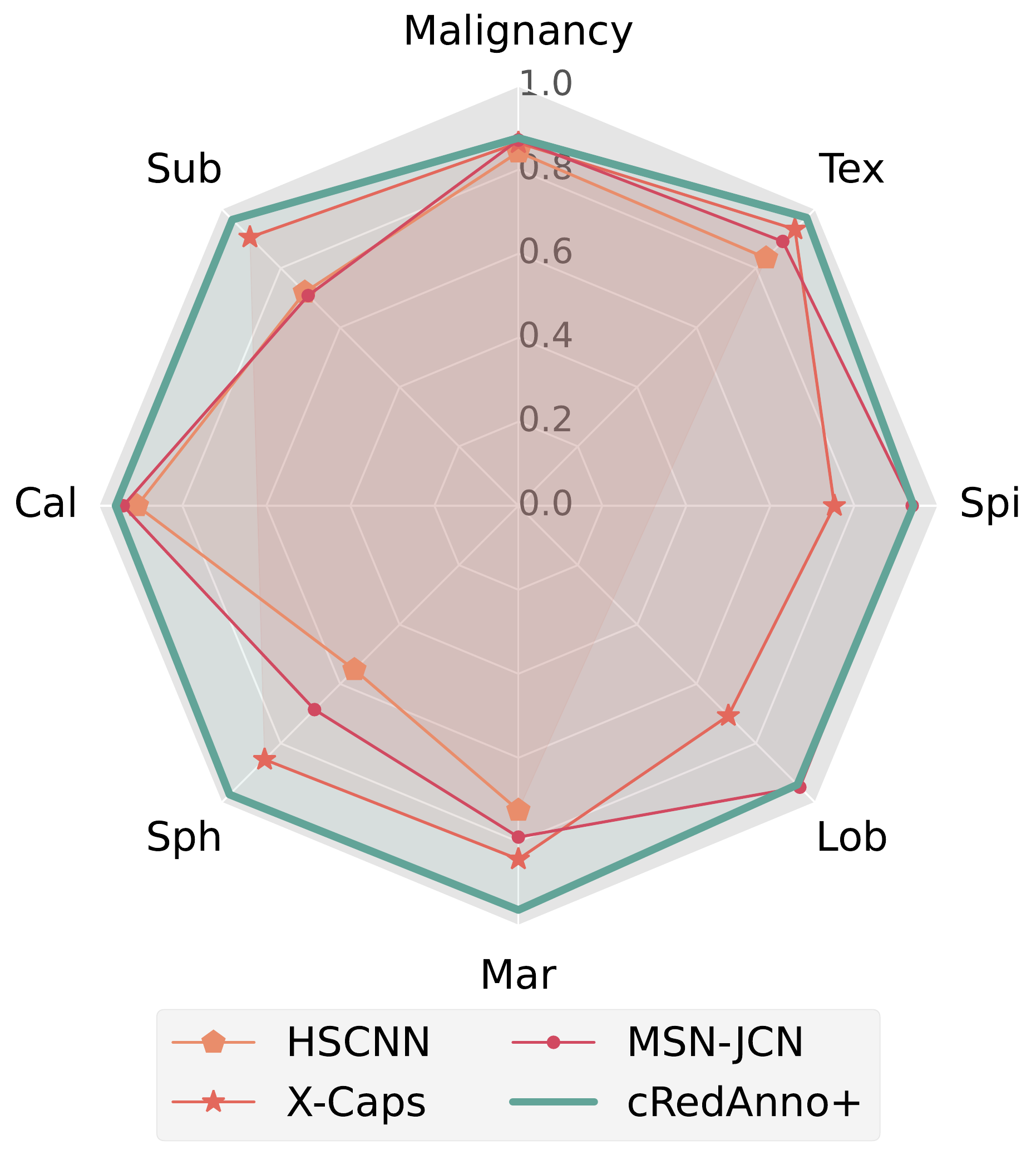}
        \label{fig:radar_a}}\hfil
    \subfloat[Partial annotation]{
        \includegraphics[width=0.47\textwidth]{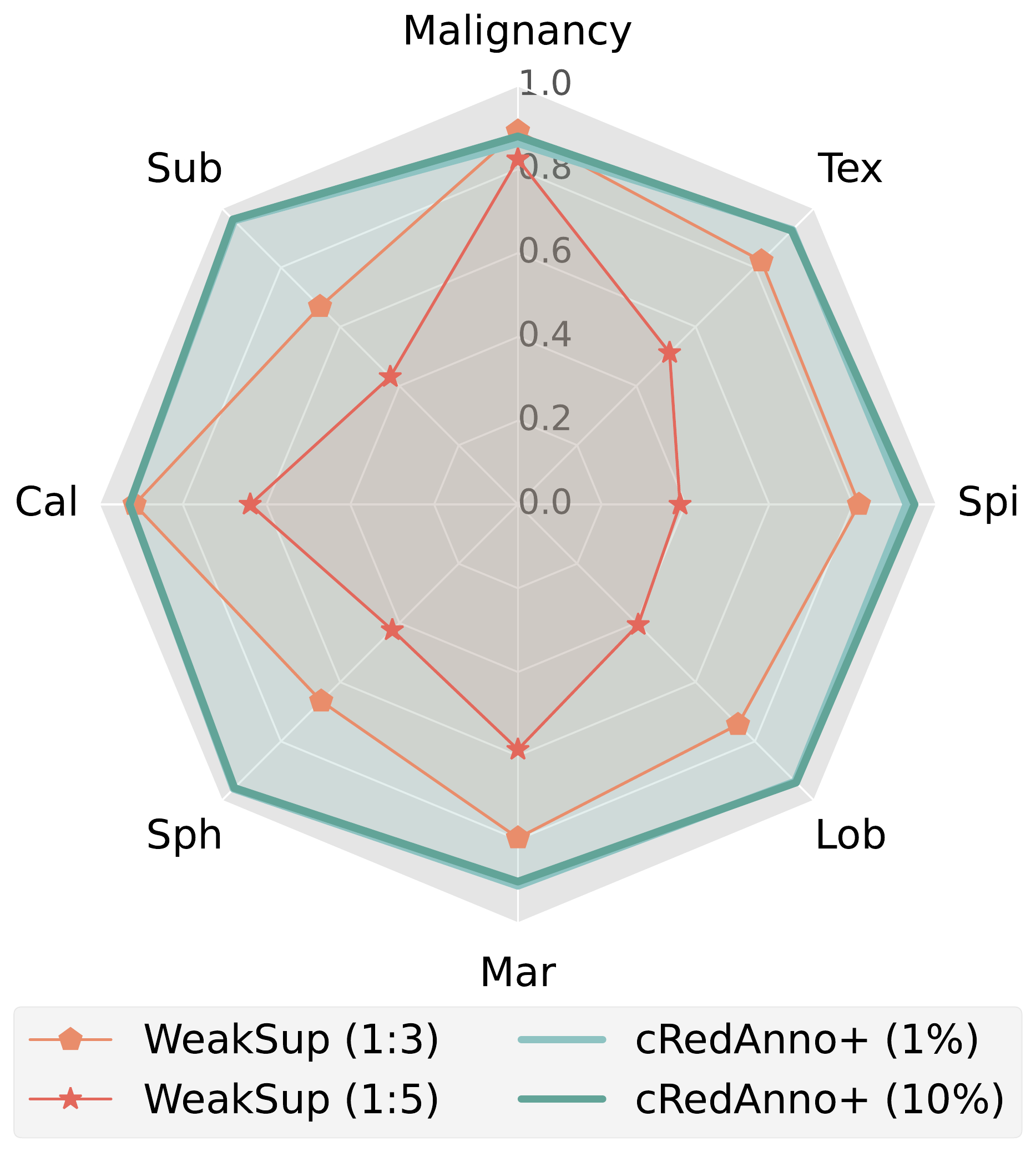}
        \label{fig:radar_b}}\\[-0.7ex]
    \caption{\textbf{Performance comparison}, in terms of prediction accuracy ($\%$) of nodule attributes and malignancy.
    Observe that \textcolor[HTML]{2F847C}{cRedAnno+} achieves simultaneously high accuracy in predicting malignancy and all nodule attributes, regardless of using either full or partial annotations.
    }
    \label{fig:radar}
\end{figure}

\subsection{Analysis of extracted features in the learned space}
\label{sec:exp_quality}

The reduction of annotations relies heavily on the separability and semantic information of the learned feature space. We use t-SNE 
to visualise the learned feature as a qualitative evaluation. Feature $f$ extracted from each testing image is mapped to a data point in 2D space. Fig.~\ref{fig:tsne_mal} to \ref{fig:tsne_tex} correspond to these data points coloured by the ground truth annotations of malignancy to nodule attribute ``texture", respectively. 

Fig.~\ref{fig:res_tsne} intuitively demonstrates the underlying correlation between malignancy and nodule attributes.
For instance, the cluster in Fig.~\ref{fig:tsne_cal} indicates that solid calcification negatively correlates with nodule malignancy. 
Similarly, Fig.~\ref{fig:tsne_lob} and Fig.~\ref{fig:tsne_spi} indicate that lobulation is associated with spiculation, both of which are positively correlated with malignancy.
These semantic correlations coincide with the radiologists' diagnostic process\cite{vlahosLungCancerScreening2018} and thereby further endorse the potential of the proposed approach as a trustworthy decision support system.

More importantly, Fig.~\ref{fig:tsne_mal} shows that even in this 2D space for visualisation, the samples show reasonably separable in both malignancy and nodule attributes. This provides the possibility to train the initial predictors using only a very small number of seed annotations, provided they are sufficiently dispersed and informative. 

\subsection{Prediction performance of nodule attributes and malignancy}
\label{sec:exp_quantity}

\begin{figure}[tbp]
    \centering
    \includegraphics[width=0.8\textwidth]{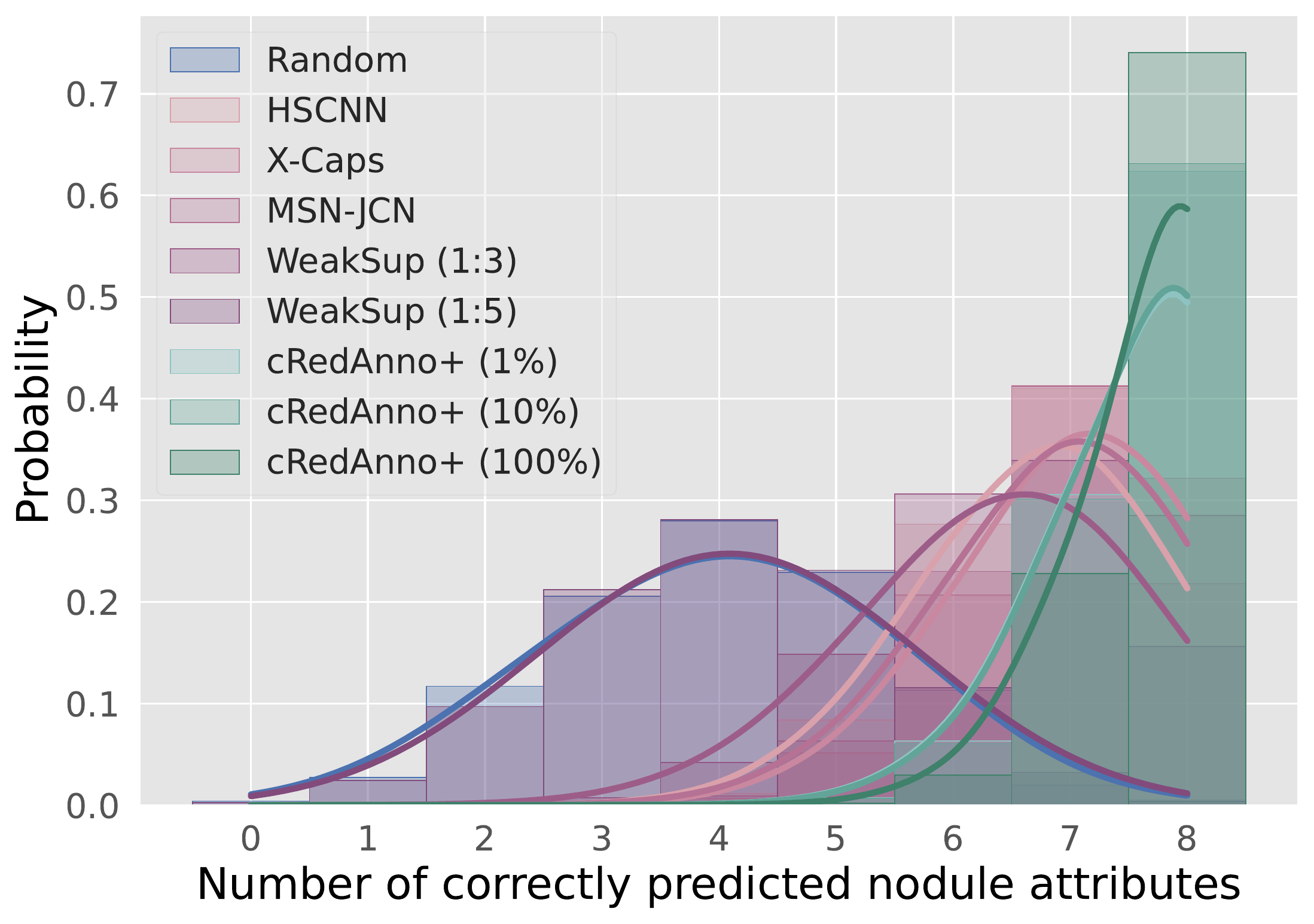}

    \caption{\textbf{Probabilities of correctly predicting a certain number of attributes for a given nodule sample.}
    Observe that \textcolor[HTML]{2F847C}{cRedAnno+} shows a more prominent probability of simultaneously predicting all $8$ nodule attributes correctly.
    }
    \vspace{-0.7ex}
    \label{fig:res_prob_ftr}
\end{figure}

The performance is evaluated quantitatively in terms of prediction accuracy of nodule attributes and malignancy. 
The evaluation procedure is the same as in the previous cRedAnno\cite{luReducingAnnotationNeed2022}:
each annotation is considered independently \cite{shenInterpretableDeepHierarchical2019}; the predictions of nodule attributes are considered correct if within $\pm1$ of aggregated radiologists' annotation \cite{lalondeEncodingVisualAttributes2020}; attribute ``internal structure" is excluded from the results because its heavily imbalanced classes are not very informative \cite{shenInterpretableDeepHierarchical2019, lalondeEncodingVisualAttributes2020, chenEndtoEndMultiTaskLearning2021, liuMultiTaskDeepModel2020, joshiLungNoduleMalignancy2021}. 

Tab.~\ref{tab:res} summarises the overall prediction performance and compares it with the state-of-the-art. 
The results show that when using only $518$ among the $730$ nodule samples and $1\%$ of their annotations for training, cRedAnno+ reaches over $90\%$ accuracy simultaneously in predicting all nodule attributes, which outperforms all previous works. 
Meanwhile, regarding nodule malignancy prediction accuracy, cRedAnno+ performs comparably with X-Caps\cite{lalondeEncodingVisualAttributes2020} and already better than HSCNN \cite{shenInterpretableDeepHierarchical2019}, which uses 3D volume data.
When using $10\%$ annotations, our malignancy prediction accuracy surpasses all other explainable competitors using full annotations, among which MSN-JCN\cite{chenEndtoEndMultiTaskLearning2021} is heavily supervised by additional information. 
Compared with the previous cRedAnno\cite{luReducingAnnotationNeed2022}, cRedAnno+ achieves comparable or even higher malignancy prediction accuracy with $10$x fewer annotations, as shown in Fig.~\ref{fig:intro_compare}.

The visualisation of the performance comparison is shown in Fig.~\ref{fig:radar}. It can be observed that our approach demonstrates simultaneously high prediction accuracy in malignancy and all nodule attributes. This substantially increases the model's trustworthiness and has not been achieved in previous works. 

In addition, we also calculate the probabilities of correctly predicting a certain number of attributes for a given nodule sample, as shown in Fig.~\ref{fig:res_prob_ftr}. The probabilities are calculated from Tab.~\ref{tab:res}. To not underestimate the performance of other compared methods, their not reported values are all assumed to be $100\%$ accuracy. 
It can be seen that cRedAnno+ demonstrates a more prominent probability of simultaneously predicting all $8$ nodule attributes correctly. The probability of correctly predicting at least $7$ attributes is higher than $90\%$, even under the extreme $1\%$ annotation condition. In contrast, WeakSup(1:5),\cite{joshiLungNoduleMalignancy2021} despite achieving $82.4\%$ accuracy in malignancy prediction, shows no significant difference in predicting nodule attributes compared to random guessing, which we consider to be the opposite of trustworthiness.

\subsection{Ablation study}
\label{sec:res_ablation}

\begin{table}[tbp]
    \vspace{-1ex}
    \centering
    \caption{\textbf{Ablation study of proposed components,} evaluated by the prediction accuracy of malignancy using $10\%$ and $1\%$ annotations. The best in each column is \textbf{bolded}. Settings of our proposed \colorbox[HTML]{E2EFD9}{cRedAnno+} and the previous \colorbox[HTML]{DFE7F3}{cRedAnno}\cite{luReducingAnnotationNeed2022} are highlighted.
    }
    \label{tab:res_ablation}
    \resizebox{\textwidth}{!}{%
    \begin{threeparttable}
        \begin{tabular}{cccccc}
        \hline
         &  &  &  & \multicolumn{2}{c}{\textbf{Maligancy accuracy}} \\ \cline{5-6} 
        \multirow{-2}{*}{\textbf{\begin{tabular}[c]{@{}c@{}}Seed sample\\ selection\end{tabular}}} & \multirow{-2}{*}{\textbf{\begin{tabular}[c]{@{}c@{}}Annotation \\ acquisition strategy\end{tabular}}} & \multirow{-2}{*}{\textbf{\begin{tabular}[c]{@{}c@{}}Pseudo \\ labelling\end{tabular}}} & \multirow{-2}{*}{\textbf{Quenching}} & \textbf{(10\%)} & \textbf{(1\%)}\tnote{$\ast$} \\ \hline
        \rowcolor[HTML]{DFE7F3} 
        random & \xmark & \xmark & \xmark & 86.65$\pm$1.39 & 80.02$\pm$8.56 \\
        random & \textbf{malignancy confidence} & \textbf{dynamic} & \textbf{\cmark} & 82.71$\pm$7.47 & 79.50$\pm$11.10 \\
        \textbf{sparse} & integrated entropy & \textbf{dynamic} & \textbf{\cmark} & 86.52$\pm$0.99 & \textbf{86.22$\pm$2.51} \\
        \textbf{sparse} & \textbf{malignancy confidence} & static & \xmark & 85.91$\pm$1.66 & 85.35$\pm$1.93 \\
        \rowcolor[HTML]{E2EFD9} 
        \textbf{sparse} & \textbf{malignancy confidence} & \textbf{dynamic} & \textbf{\cmark} & \textbf{87.86$\pm$1.99} & \textbf{86.22$\pm$2.51} \\ \hline
        \end{tabular}%
        \begin{tablenotes}
            \item[$\ast$] Does not contain requested annotations.
        \end{tablenotes}
    \end{threeparttable}
    }
    \vspace{-0.7ex}
\end{table}

We validate the proposed annotation exploitation mechanism by ablating each component as a row shown in Tab.~\ref{tab:res_ablation}. The standard deviation when using $1\%$ annotations shows that sparse seeding plays a crucial role in stabilising performance. The sum entropy\cite{settlesUncertaintySampling2012} integrating malignancy and all nodule attributes was also experimented with as an alternative acquisition strategy, but exhibits impaired prediction accuracy. Quenching, which enables dynamic pseudo labelling, also proves necessary for the boosted performance.

\section{Conclusion}
\label{sec:conclustion}

In this paper, we propose cRedAnno+ to improve the prediction accuracy and robustness of the previous work in self-explanatory lung nodule diagnosis. 
Our experiments show that the proposed annotation exploitation mechanism enables comparable or even higher accuracy with 10x fewer annotations, meanwhile being more robust. 
Furthermore, cRedAnno+ is the first to reach over $90\%$ accuracy simultaneously in predicting all nodule attributes with only hundreds of samples and $1\%$ of their annotations, which adds significantly to the trustworthiness. 
The limitations of this work remain in its simple predictor architecture and not carefully-tuned hyperparameters, and its generalisability is yet to be verified on other suitable datasets.

\vfill
\pagebreak

\section{Acknowledgements}
\label{sec:acknowledgments}

No funding was received for conducting this study. The authors have no relevant financial or non-financial interests to disclose.

\section{Compliance with Ethical Standards}

This research study was conducted using the open-access LIDC dataset\cite{armatoLungImageDatabase2011}. Ethical approval was not required as confirmed by the license attached with the open-access data.


\begin{spacing}{0.9}
\bibliographystyle{IEEEbib}
\bibliography{Explainability}
\end{spacing}

\end{document}